\documentclass[sigplan, nonacm]{acmart}
\renewcommand\footnotetextcopyrightpermission[1]{}
\makeatletter
\renewcommand{\@authorfont}{\large\normalfont}
\renewcommand{\@affiliationfont}{\normalsize\normalfont}
\patchcmd{\@mkauthors@iii}
  {\mathchardef\UrlBreakPenalty=10000\nolinkurl}
  {\mathchardef\UrlBreakPenalty=10000\small\nolinkurl}
  {}{\ClassWarning{acmart}{Could not reduce email font size}}
\patchcmd{\@mkauthors@iii}
  {\par\bgroup\mathchardef\UrlBreakPenalty=10000\nolinkurl}
  {\par\bgroup\mathchardef\UrlBreakPenalty=10000\small\nolinkurl}
  {}{\ClassWarning{acmart}{Could not reduce email font size}}
\makeatother

\AtBeginDocument{%
  }

\usepackage[a-1b]{pdfx}

\usepackage{amsmath, amsfonts}
\usepackage{amssymb}
\usepackage{graphicx}
\usepackage{textcomp}
\usepackage[dvipsnames]{xcolor}
\usepackage{subcaption}
\usepackage{xspace}
\usepackage{tikz} % to draw colored boxes
\usepackage{bbding} % for checkmark and xsolid 
\usepackage{diagbox} % to draw slash in tables
\usepackage[normalem]{ulem}
\usepackage{xcolor}

\usepackage{algorithm}
\usepackage{algpseudocodex}
\usepackage{algorithmicx}

\usepackage{environ}
\usepackage{adjustbox}

\usepackage{pgfplots}
\usetikzlibrary{patterns}
\usepgfplotslibrary{groupplots}

\usepackage{todonotes}

\usepackage{enumitem}

\usepackage[most]{tcolorbox}

\usepackage{wrapfig}

\usepackage{amsthm}  

\usetikzlibrary{spy}

\usepackage{tabularx}
\usepackage{soul}

\usepackage{multirow}

\usepackage{capt-of}

\usepackage{mathpartir}
\usepackage{mathtools}

\usepackage{marginnote}

\usepackage{balance}

\newcommand{\ourtool}{Elle$^{+}$\xspace}

\newcommand{\inlsec}[1]{\smallskip\noindent\textbf{#1.}}

\newcommand{\mylist}[1]{\textcolor{black}{\texttt{#1}}}

\newcommand{\ser}{\textsc{Serializability}\xspace}

\definecolor{ps1green}{HTML}{B7E6A5} % green series 1-7, greater number leads to a the thicker color
\definecolor{ps2green}{HTML}{7CCBA2}
\definecolor{ps3green}{HTML}{46AEA0}
\definecolor{ps4green}{HTML}{089099}
\definecolor{ps5green}{HTML}{00718B}
\definecolor{ps6green}{HTML}{045275}
\definecolor{ps7green}{HTML}{003147}

\definecolor{diverge1}{HTML}{045275}
\definecolor{diverge2}{HTML}{089099}
\definecolor{diverge3}{HTML}{EF8733}
\definecolor{diverge4}{HTML}{FCDE9C}
\definecolor{diverge5}{HTML}{F0746E}
\definecolor{diverge6}{HTML}{DC3977}
\definecolor{diverge7}{HTML}{7C1D6F}

\definecolor{equalsoneoptcolor}{RGB}{253, 243, 208}
\definecolor{bigandsmallcyclesoptcolor}{RGB}{217, 232, 214}
\definecolor{otheroptcolor}{RGB}{251, 231, 207}
\newcommand{\true}{\textsf{true}}
\newcommand{\false}{\textsf{false}}

\newcommand{\set}[1]{\{#1\}}

\newcommand{\emptypost}{\mathit{empty}}

\newcommand{\uniqueval}{unique-value}

\newcommand{\duplicateval}{duplicate-value}

\newcommand{\readevent}{{\sf R}}

\newcommand{\appendevent}{{\sf A}}

\newcommand{\gar}{{\sf gar}}
\newcommand{\gvis}{{\sf gvis}}

\newcommand{\T}{\mathcal{T}}
\providecommand{\G}{}
\renewcommand{\G}{\mathcal{G}}
\renewcommand{\H}{\mathcal{H}}

\newcommand{\operations}{\mathcal{O}}

\newcommand{\SO}{\textsf{\textup{SO}}}
\newcommand{\WR}{\textsf{\textup{WR}}}
\newcommand{\WW}{\textsf{\textup{WW}}}
\newcommand{\RW}{\textsf{\textup{RW}}}

\providecommand{\so}{}
\renewcommand{\so}{\textsf{\textup{so}}}
\providecommand{\wr}{}
\renewcommand{\wr}{\textsf{\textup{wr}}}
\newcommand{\ww}{\textsf{\textup{ww}}}
\newcommand{\rw}{\textsf{\textup{rw}}}
\newcommand{\app}{\textsf{\textup{ao}}}

\renewcommand{\ae}{\mathcal{A}}

\newcommand{\rel}[1]{\xrightarrow{#1}}
\newcommand{\relimm}[1]{\xrightarrow[\text{imm}]{#1}}

\newcommand{\po}{{\sf po}}

\newcommand{\vis}{\textsc{vis}}
\newcommand{\ar}{\textsc{ar}}
\newcommand{\intaxiom}{\textsc{Int}}

\newcommand{\intextaxiom}{\textsc{IntExt}}

\newcommand{\sessionaxiom}{\textsc{Session}}

\newcommand{\totalvis}{\textsc{TotalVis}}
\newcommand{\C}{\mathcal{C}}

\newcommand{\wwopcons}{\C^{\ww}}
\newcommand{\wropcons}{\C^{\wr}}

\makeatletter
\newsavebox{\measure@tikzpicture}
\NewEnviron{scaletikzpicturetowidth}[1]{%
  \def\tikz@width{#1}%
  \def\tikzscale{1}\begin{lrbox}{\measure@tikzpicture}%
  \BODY
  \end{lrbox}%
  \pgfmathparse{#1/\wd\measure@tikzpicture}%
  \edef\tikzscale{\pgfmathresult}%
  \BODY
}
\makeatother

\colorlet{sgray}{gray!80!black}
\colorlet{sgreen}{green!40!black}
\colorlet{sred}{red!70!black}
\colorlet{sbrown}{brown!80!black}
\colorlet{sblue}{blue!50!black}

\newenvironment{proofsketch}{%
\renewcommand{\proofname}{Proof Sketch}\proof}{\endproof}
\begin{document}

\title{Extending Elle for Transaction Workloads with\\  Duplicate Values}
\titlenote{This work was accepted by and will be presented at the 1st Symposium on Consistency Checking Principles (SCCP 2026).}
\author{Zhiheng Cai}
\affiliation{%
  \institution{Tsinghua University}
  \country{}
}
\email{cai-zh24@mails.tsinghua.edu.cn}

\author{Si Liu}
\affiliation{%
  \institution{Texas A\&M University}
  \country{}
}
\email{si.liu@tamu.edu}

\author{Hengfeng Wei}
\affiliation{%
  \institution{Hunan University}
  \country{}
}
\email{hfwei@hnu.edu.cn}

\author{Yuxing Chen}
\affiliation{%
  \institution{Renmin University of China}
  \country{}
}
\email{axinggu@gmail.com}



% abs.tex 

\begin{abstract}
Elle is one of the most widely adopted black-box isolation validators. 
 It crucially relies on the unique-value assumption
  for sound and efficient isolation validation.
Yet,
transaction workloads with duplicate values are highly relevant in practice: they naturally arise in real database systems, and many isolation bugs manifest only in their presence. 
In this paper, we extend Elle to handle such workloads by introducing a fine-grained dependency model
that enables reasoning about dependencies 
between individual operations.
We  establish the soundness and completeness 
of our approach and implement it in a prototype.
We also demonstrate its effectiveness in
 detecting isolation bugs and its promising performance.

%    Using \ourtool, we successfully reproduce two known serializability bugs 
   % that can only be triggered in the presence of duplicate values.
   %  Our evaluation further demonstrates promising performance, suggesting the practical viability of the approach.
	
%Elle, the core checker of the Jepsen testing framework, 
%checks list-append histories 
%by inferring $\WR$ and $\WW$ dependencies from read lists. 
%This inference relies on the \uniqueval{} assumption: 
%each appended value uniquely identifies the operation that produced it. 
%Without this assumption, 
%Elle's inference can be disrupted. 
%However, recent bug reports reveal that 
%some bugs may only manifest under duplicate write values, 
%which is beyond Elle's current applicability. 
%
%In this paper, we extend Elle to handle duplicate-value list-append histories. 
%We propose an operation-level modeling for these histories.
%The key idea is to unfold each read list into its prefixes, 
%thereby exposing the step-by-step evolution of the list. 
%This modeling enables us to search for 
%a valid serializable explanation of the observed lists despite
%the ambiguity caused by duplicate values. 
%Our prototype implementation demonstrates the effectiveness of our approach
%by reproducing known bugs in existing database systems.
\end{abstract}





\maketitle

% intro.tex

\section{Introduction}

% \nobi{emphasize the application scenario of elle/jepsen: testing (bugs only manifest when duplicate values are present~\cite{x,y}). (We might also consider another scenario: a cloud database system auditor with limited control over workload/test generation, but it seems a bit odd with jepsen which is a TESTER...)}

% \nobi{emphasize soundness (or completeness?). duplicate values break Elle's  guarantee}

% \nobi{a small paragraph on related work suffices in Intro}

% \textcolor{NavyBlue}{
%   \begin{itemize}
%     \item verifying database consistency is very hard.
%     \item Jepsen is effective in finding bugs. the workflow of jepsen.
%     \item Elle, the efficiency checker, Adya style's dependency graph, \uniqueval, list-append histories to detect bugs.
%     \item Duplicate values may unveil new bugs. we want to extend Elle to \duplicateval{} histories.  
%     \item challenges: $\WW$ and $\WR$ uncertainty. 
%     \item our solution: operation-level modeling and append-order recovery. 
%     \item evaluation (can reproduce \duplicateval{} bugs), conclusion. 
%   \end{itemize}
% }

Ensuring that   database systems 
uphold their promised isolation guarantees,
e.g., serializability,
 has long been a central concern
in both research and practice. 
 Violating these guarantees can lead to undesirable data anomalies, e.g., lost updates, as well as severe security and reliability issues~\cite{DBLP:conf/sigmod/WarszawskiB17,10.1145/3597503.3639207}.

Black-box validation~\cite{Complexity:OOPSLA2019,Cobra:OSDI2020,Elle:VLDB2020,PolySI:VLDB2023,Plume:OOPSLA2024,veristrong,AWDIT:PLDI2025,Viper:EuroSys2023,MiniTransactions:ICDE2025,leopard} 
offers an effective practical approach to addressing this concern, as
 modern database systems often have large  codebases that are inaccessible or too difficult to verify directly.
In this approach, a validator collects execution histories of database transactions as an external observer 
 and \emph{verifies} whether these histories satisfy the isolation level in question,
  i.e., without producing false positives or missing bugs.

Elle~\cite{Elle:VLDB2020} stands out as one of the most widely adopted black-box isolation validators. 
The reasons are mainly twofold. 
First, although the
\emph{history verification problem} is NP-complete in general~\cite{Complexity:OOPSLA2019}, Elle achieves near-linear validation time in practice, allowing it to scale to real-world workloads. Second, as part of the Jepsen framework~\cite{Jepsen}, Elle can leverage techniques such as fault injection to stress-test database systems and explore a much broader space of system behaviors.
%Together, they have uncovered numerous bugs across  over two dozen database systems to date~\cite{Jepsen-Analyses}.

Key to Elle's
effectiveness and efficiency 
is the co-design of 
a \emph{list-append} workload generator and
 a  transactional dependency inference mechanism
 based on Adya's theory~\cite{Adya:PhDThesis1999}.
For example, a read returning the list 
\mylist{[1,2]}
allows Elle to infer the version order
(or the $\WW$ dependency)
 between writes: 
the append that wrote 
\mylist{1}
 must precede the append that wrote  \mylist{2}
in the database.
 In contrast, most existing  isolation validators 
 are  built on the traditional read-write register setting,
where they must 
``guess'' all possible
$\WW$ dependencies, 
a key source of the problem's computational complexity.

Crucially,
Elle relies on the \emph{unique-value} assumption:
each  value read can be deterministically matched to a single write. 
Otherwise,  in the presence of \emph{duplicate}    values, 
a read returning a list such as 
\mylist{[1,2,1]}
 no longer allows Elle to reliably
determine which append of 
\mylist{1}
  produced each observed value (i.e., the $\WR$ dependency), 
thereby preventing accurate inference of the $\WW$ dependency between the two appends.

Yet, 
transaction workloads with duplicate values
 are highly relevant in practice. 
 Recent findings~\cite{MySQL-Bug,MariaDB-Bug} 
reveal that 
many isolation bugs arise \emph{only}
 in the presence of duplicate   values. 
Moreover, 
duplicate-value workloads 
 can 
 \emph{naturally} occur in real   systems,
e.g., due to retry mechanisms, and may expose subtle isolation bugs~\cite{Elle:VLDB2020}.
Thus,  checkers
 relying solely on unique-value workloads may fail to detect these bugs.
Due to the  randomized nature of black-box testing, 
rediscovering even a single missed bug may require substantial effort.

In this paper, we extend Elle to support database workloads with duplicate values
 while preserving 
soundness,  
 completeness, 
 and
   much of its practical efficiency.

One may wonder
whether handling duplicate values
 merely requires reasoning over additional uncertain
  $\WR$ dependencies,
  as in the recent solution for the 
  read-write register setting~\cite{veristrong}.
For example, one may attempt   possible matchings 
between the appends of 
\mylist{1}
 and 
the corresponding values read in the list, 
and then check whether any resulting Adya dependency graph is acyclic,
 which is equivalent to satisfying serializability.
Yet, 
this may yield a spurious serial interpretation of the underlying execution,
which in turn may cause genuine bugs to be missed
(see Section~\ref{sec:solution} for details).

We address this challenge 
by introducing a fine-grained dependency model
 that enables reasoning about dependencies 
 at the 
 \emph{operation level}
  rather than the transaction level.
Our approach
unfolds each 
returned
list  into its prefixes,
thereby recovering
operation-level $\WR$ dependencies between append and read operations.
  These dependencies, in turn, 
enable precise inference of $\WW$ dependencies 
despite duplicate values.
We formally establish the soundness and completeness of our approach.
  
%\orange{
%The key observation is that 
%a returned list records a sequence of successive extensions of the same list. 
%Each extension must be explained by a source append operation, 
%and consecutive extensions impose ordering constraints on the corresponding appends. 
%Rather than transaction-level dependencies, 
%our model explicitly accounts for dependencies 
%between individual operations. 
%Our approach unfolds each list read into its prefixes, 
%thereby recovering fine-grained operation-level $\WR$ dependencies 
%between appends and reads. 
%These dependencies in turn allow us to derive precise $\WW$ dependencies 
%even in the presence of duplicate values.
%}
 
We develop an SMT-based prototype, \ourtool, 
that uses hyper-polygraphs~\cite{veristrong} to encode dependency constraints.
Our current implementation focuses on checking serializability.
Using \ourtool, we successfully  reproduced two known serializability bugs in MySQL and MariaDB, 
both of which can only be triggered in the presence of duplicate values and 
  are thus beyond the scope of Elle.
We further show that 
 \ourtool   achieves promising performance, 
suggesting its practical viability.

\begin{figure*}[h!]
% \begin{figure*}[t]
	\centering
	\begin{subfigure}[b]{0.22\textwidth}
		\centering
		\includegraphics[width=\textwidth]{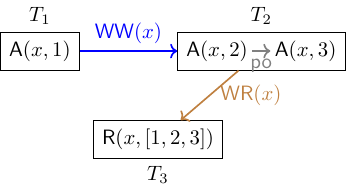}
			\captionsetup{skip=5pt}
		\caption{A \uniqueval{} list-append history.}
		\label{fig:list-uv}
	\end{subfigure}
	\hfill
	\begin{subfigure}[b]{0.21\textwidth}
		\centering
		\includegraphics[width=\textwidth]{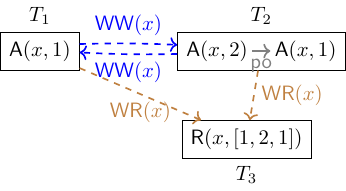}
					\captionsetup{skip=5pt}
		\caption{A \duplicateval{} history with constraints.}
		\label{fig:list-dup-constraints}
	\end{subfigure}
	\hfill
	\begin{subfigure}[b]{0.21\textwidth}
		\centering
		\includegraphics[width=\textwidth]{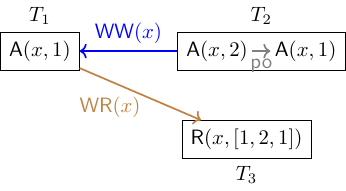}
					\captionsetup{skip=5pt}
		\caption{An unfaithful acyclic resolution.}
		\label{fig:list-dup-unfaithful}
	\end{subfigure}
	\hfill
	\begin{subfigure}[b]{0.24\textwidth}
		\centering
		\includegraphics[width=\textwidth]{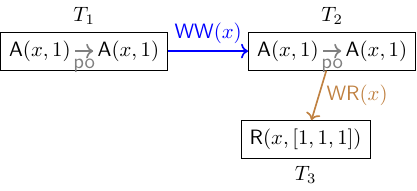}
					\captionsetup{skip=5pt}
		\caption{ A false negative caused by  transaction-level modeling.}
		\label{fig:list-dup-false-negative}
	\end{subfigure}
				\captionsetup{skip=8pt}
	\caption{Example histories showing that transaction-level dependency modeling is sufficient for \uniqueval{} list-append histories but insufficient when duplicate values are present. }
	\label{fig:list-motivating}
	\vspace{-3ex}
\end{figure*}

\section{How Elle Infers  Dependencies}

%\zhiheng{note the lower-case ``unique-value'' and ``duplicate-value'' in section titles}

Elle~\cite{Elle:VLDB2020} considers a distributed key-value store over a set of keys
(denoted by $x, y, z, \dots$), where each key identifies a list.
A list-append history 
consists of 
two kinds of operations:
an append operation $\appendevent(x, v)$ appending value $v$ to the list
stored at key $x$ and a read operation $\readevent(x, \ell)$ reading
$x$ and returning a list $\ell$ of values. 
The \uniqueval{} assumption requires that, 
for each key $x$, each value $v$ is appended at most once. 

A transaction is a finite set of operations 
totally ordered by the \emph{program order} $\po$. 
A history $\H$ is a pair $(\T, \SO)$, 
where $\T$ is the set of transactions and 
$\SO$ is the session order.
A list-append history satisfies serializability
 iff 
its transactions can be totally ordered in a way that 
respects the session order, 
such that each read operation observes exactly the list 
produced by all preceding appends to the
same key. 

Elle checks  a \uniqueval{} list-append history
 by 
 first
constructing an Adya dependency graph~\cite{Adya:PhDThesis1999}.
The graph contains three kinds of transaction-level dependencies: 
$\WR$ relates a writer to a reader of its value,
$\WW$ orders transactions that write the same key,
and $\RW$ relates a reader of a version to a later writer of that key.
The history then satisfies 
serializability
 iff $\SO \cup \WR \cup \WW \cup \RW$ is acyclic.

Figure~\ref{fig:list-uv} shows an example \uniqueval{} list-append history.  
%We use this history to illustrate how Elle infers $\WR$ and $\WW$ dependencies, 
%constructs dependency graphs and checks serializability.    
Consider the list   \mylist{[1,2,3]}  read by $T_3$.
The
 last value, \mylist{3}, uniquely identifies the $\WR$ dependency
$T_2 \rel{\WR(x)} T_3$,
while
the relative positions of \mylist{1}  and \mylist{2} imply the $\WW$ dependency
$T_1 \rel{\WW(x)} T_2$.
%The  dotted arrows
 %indicate the append operations
%responsible for each value read by $T_3$. 
Therefore,
 in the \uniqueval{} setting, 
the list itself contains sufficient information to recover these dependencies. 
The resulting dependency graph is acyclic and admits the serial order
$T_1 \rightarrow T_2 \rightarrow T_3$.
%Elle then derives $\RW$ dependencies from the 
%recovered
% $\WR$ and $\WW$ dependencies 
% and
%checks whether the resulting dependency graph is acyclic.

% solution.tex

\section{Extending Elle for Duplicate Values } 
\label{sec:solution}

%\subsection{Dependency Ambiguity under Duplicate Values} 
In the presence of duplicate values,
Elle can no longer directly recover $\WR$ and $\WW$ dependencies from read lists. 
Figure~\ref{fig:list-dup-constraints} 
shows an example, where
 both $T_1$ and $T_2$ append value \mylist{1} to key $x$.
As a result, when $T_3$ reads \mylist{[1,2,1]},
the two occurrences of \mylist{1} cannot be uniquely mapped to their source appends, 
  and different assignments may induce different $\WR$ dependencies.
Such uncertainties also extend to $\WW$ dependencies, 
as the same read list may imply different, or even conflicting, orderings of appends.

\subsection{A Strawman Using Transactional Dependencies}
A natural attempt to address 
this
 issue is to represent each 
 $\WR$ or $\WW$ uncertainty
  as a choice set 
 and ask whether some resolution yields an acyclic dependency graph.
% This is essentially the idea used for resolving uncertain $\WW$ dependencies in the read-write register setting.
 
Figure~\ref{fig:list-dup-constraints} illustrates this idea.
As both $T_1$ and $T_2$ write to
 key $x$, 
 we introduce the choice set
$\set{T_1 \rel{\WW(x)} T_2, 
	T_2 \rel{\WW(x)} T_1}$.
Likewise, as the final value read by $T_3$ may originate from either transaction, we introduce
$\set{T_1 \rel{\WR(x)} T_3, T_2 \rel{\WR(x)} T_3}$.
Under this 
transaction-level  
modeling, a history is deemed serializable if one edge can be selected from each choice set such that the resulting dependency graph is acyclic.

However, this transaction-level modeling is insufficient.
Figure~\ref{fig:list-dup-unfaithful} shows an acyclic resolution
$T_2 \rel{\WW(x)} T_1 \rel{\WR(x)} T_3$.
Despite being acyclic, it cannot account for the observed read list
\mylist{[1,2,1]}.
In particular, the resolution overlooks an append by $T_2$ that must occur between the append of 
\mylist{2} by $T_2$ and the append of 
\mylist{1}
 by $T_1$.
Consequently,
it implies a
 different evolution 
 of the list
  than the one actually observed.

Even worse,
 such unfaithful resolutions can 
 mask isolation bugs.
Figure~\ref{fig:list-dup-false-negative} shows a non-serializable history,
 in which both $T_1$ and $T_2$ append  \mylist{1} twice, while $T_3$ reads \mylist{[1,1,1]}.
No serial execution can produce this observation: 
by atomicity, 
$T_3$ can observe either zero, two, or four occurrences of \mylist{1}, 
but never three.
Yet,
this approach
admits an acyclic resolution.

\begin{figure}[t]
	\centering
	\begin{subfigure}[b]{0.45\textwidth}
		\centering
		\includegraphics[width=0.9\textwidth]{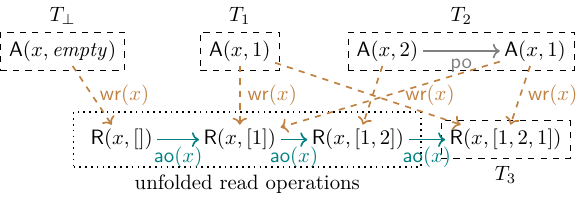}
		\captionsetup{skip=5pt}
		\caption{Unfolded reads and operation-level constraints
			for the history in Figure~\ref{fig:list-dup-constraints}.
			The $\ww$ constraints are omitted for readability.}
		\label{fig:list-dup-op-constraints}
	\end{subfigure}
	\begin{subfigure}[b]{0.45\textwidth}
		\centering
		\includegraphics[width=0.9\columnwidth]{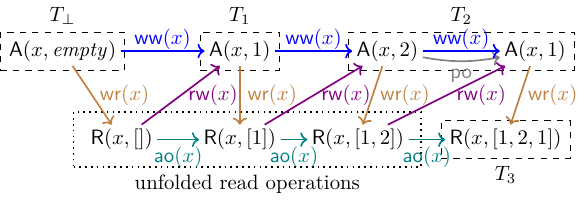}
		\captionsetup{skip=5pt}
		\caption{
			A valid resolution witnessing serializability at the operation level. 
			The $\so$ edges from $T_{\bot}$ to all other transactions are omitted for readability. } 
		\label{fig:list-dup-solution}
	\end{subfigure}
\begin{subfigure}[b]{0.45\textwidth}
	\centering
  \includegraphics[width=0.9\columnwidth]{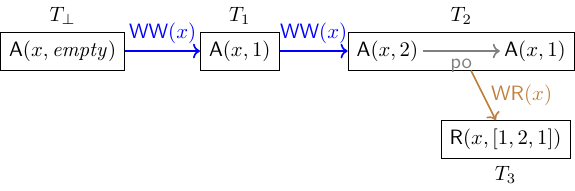}
	\captionsetup{skip=5pt}
  \caption{The corresponding  transaction-level dependency graph for the valid resolution.}
\label{fig:list-dup-solution-txn}
\end{subfigure}
	\captionsetup{skip=8pt}
	\caption{Illustrating list unfolding and the resulting operation- and transaction-level dependency graphs for the duplicate-value history in Figure~\ref{fig:list-dup-constraints}. }
	\label{fig:list-solution}
	\vspace{-3ex}
\end{figure}

\vspace{-2ex}
\subsection{Our Solution:     List Unfolding} 
The limitations of the strawman approach point to
  the need for a faithful characterization of dependencies that:
  (i) identifies the source append responsible for each successive list extension, 
   rather than only the transaction producing the final value; 
    and
  (ii) ensures that consecutive values in a list correspond to consecutive appends in the $\WW$ order, 
  with no intervening append on the same key.
  
To this end, 
 we shift \emph{from transaction-level to operation-level} reasoning.
The key idea is 
  to unfold
  each read list into its successive prefixes,  
  thereby identifying the append responsible for each successive extension of the observed list.

Specifically, for each read 
$r \triangleq \readevent(x, [v_1, v_2, \cdots, v_n])$, we
introduce $n$ auxiliary reads
$r'_0, r'_1, \cdots, r'_{n-1}$, 
where $r'_i$ reads the length-$i$ prefix
$[v_1,\ldots,v_i]$ of the original list
($r’_0$ reads the empty list, while $r$ itself reads the full list). 
To capture these successive list extensions, 
 we introduce a new order $\app$, called the \emph{append order}.
 For every read
 and its unfolded reads, 
 we then add the chain of $\app(x)$-labeled edges:
 $r'_0 \rel{\app(x)} r'_1
 \rel{\app(x)} \cdots
 \rel{\app(x)} r'_{n-1}
 \rel{\app(x)} r$.
%  \small
%  \[
%  r'_0 \rel{\app(x)} r'_1
%  \rel{\app(x)} \cdots
%  \rel{\app(x)} r'_{n-1}
%  \rel{\app(x)} r.
%  \]
%  \normalsize

%\[
 % r'_i \triangleq \readevent(x, [v_1, v_2, \cdots, v_i])
 % \text{ for } 0 \leq i < n,
%\]
%with the convention that $[v_1, \cdots, v_0] = []$.
%That is, 
%$r'_0$ reads the empty list, $r'_1$ reads the prefix $[v_1]$,
%and so on, while the original read $r$ reads the full list
%$[v_1, \cdots, v_n]$.
%each auxiliary read corresponds to a prefix of the original read list, 
% while $r$ itself reads the full list $[v_1, \cdots, v_n]$.

%To record that these reads correspond to successive extensions of the
%same list, we introduce a new edge label $\app$, called the
%\emph{append order}. For every read
%$r \triangleq \readevent(x, [v_1, v_2, \cdots, v_n])$ and its unfolded
%reads $r'_0, r'_1, \cdots, r'_{n-1}$, we add exactly the following
%$\app(x)$-labeled edges:
%\[
%  r'_0 \rel{\app(x)} r'_1
%  \rel{\app(x)} \cdots
%  \rel{\app(x)} r'_{n-1}
%  \rel{\app(x)} r .
%\]

Figure~\ref{fig:list-dup-op-constraints}
depicts 
how our approach handles the duplicate-value history in Figure~\ref{fig:list-dup-constraints}.
By unfolding the read list of $T_3$,
uncertain $\WR$ dependencies are captured by the following operation-level choice sets:\footnote{Following prior work~\cite{Complexity:OOPSLA2019,veristrong,PolySI:VLDB2023,Plume:OOPSLA2024},  we additionally introduce a synthetic initial transaction $T_{\bot}$, which initializes each list to $\emptypost$.}
% $\set{T_{\bot}\!:\!\appendevent(x, \emptypost) \rel{\wr(x)} \readevent(x, [])}, 
%  \set{T_1\!:\!\appendevent(x, 1) \rel{\wr(x)} \readevent(x, [1]),
% 		T_2\!:\!\appendevent(x, 1) \rel{\wr(x)} \readevent(x, [1])},  
%  \set{T_2\!:\!\appendevent(x, 2) \rel{\wr(x)} \readevent(x, [1, 2])}, 
%  \set{T_1\!:\!\appendevent(x, 1) \rel{\wr(x)} \readevent(x, [1, 2, 1]),
% 		T_2\!:\!\appendevent(x, 1) \rel{\wr(x)} \readevent(x, [1, 2, 1])}.
% $
\vspace{-1ex}
\small
\[
\begin{aligned}
	& \set{T_{\bot}\!:\!\appendevent(x, \emptypost) \rel{\wr(x)} \readevent(x, [])}, \\[-3pt]
	& \set{T_1\!:\!\appendevent(x, 1) \rel{\wr(x)} \readevent(x, [1]),
		T_2\!:\!\appendevent(x, 1) \rel{\wr(x)} \readevent(x, [1])}, \\[-3pt]
	& \set{T_2\!:\!\appendevent(x, 2) \rel{\wr(x)} \readevent(x, [1, 2])}, \\[-3pt]
	& \set{T_1\!:\!\appendevent(x, 1) \rel{\wr(x)} \readevent(x, [1, 2, 1]),
		T_2\!:\!\appendevent(x, 1) \rel{\wr(x)} \readevent(x, [1, 2, 1])}.
\end{aligned}
\]
\normalsize
Here, 
 $T\!:\!\appendevent(\_, \_)$ 
 denotes an append operation in transaction $T$, 
and 
(lower-case) $\wr$ 
relations denote read-from 
dependencies between individual operations.

%We additionally introduce a synthetic initial transaction $T_{\bot}$,
%which provides a distinguished initial value $\emptypost$ for each list.
%Based on these unfolded operations, 
%we can express dependencies at the operation level.

%As shown in Figure~\ref{fig:list-dup-op-constraints}, the possible
%$\WR$ dependencies are captured by the following operation-level
%choice sets:

Similarly, for $\WW$ dependencies, 
we construct a choice set for each pair of append operations on the same key:
\small
\[
\begin{aligned}
  &\set{T_1\!:\!\appendevent(x, 1) \rel{\ww(x)} T_2\!:\!\appendevent(x, 2),
        T_2\!:\!\appendevent(x, 2) \rel{\ww(x)} T_1\!:\!\appendevent(x, 1)}, \\[-3pt]
  &\set{T_1\!:\!\appendevent(x, 1) \rel{\ww(x)} T_2\!:\!\appendevent(x, 1),
        T_2\!:\!\appendevent(x, 1) \rel{\ww(x)} T_1\!:\!\appendevent(x, 1)}, \\[-3pt]
  &\set{T_2\!:\!\appendevent(x, 2) \rel{\ww(x)} T_2\!:\!\appendevent(x, 1),
        T_2\!:\!\appendevent(x, 1) \rel{\ww(x)} T_2\!:\!\appendevent(x, 2)}.
\end{aligned}
\]
\normalsize
We omit the $\WW$ choices involving the initial append operation, 
since it is known to precede all other appends.
%these choices are likewise omitted from Figure~\ref{fig:list-dup-op-constraints} for readability.

This operation-level dependency model
 enables us to identify the source append operation responsible for each successive extension of a  list.
Once the operation-level $\wr$ dependencies have been selected,
 we further constrain the operation-level $\ww$ order.

 Specifically, for every pair of (unfolded) reads $r$ and $r'$ such that
 $r \rel{\app(x)} r'$,
 if the selected $\wr$ dependencies are
 $w \rel{\wr(x)} r$ 
 and
 $w’ \rel{\wr(x)} r'$,
 then the resolution must also select
 $w \rel{\ww(x)} w'$. 
 Furthermore, $w$ and $w'$ must be adjacent in the $\ww(x)$ order: there must be no append operation $w''$ on key $x$ such that
 $w \rel{\ww(x)} w''  \land w'' \rel{\ww(x)} w'$.

This condition,
 which we call the \emph{append-order condition}, 
 enforces the list semantics that adjacent prefixes of a read list correspond to adjacent append operations.

%This condition captures the list semantics: 
%two adjacent prefixes in a read list must be produced 
%by two adjacent append operations.  

In addition,
 a valid resolution
  witnessing serializability 
must  be acyclic at
 \emph{both}
 the operation and transaction levels. 

First, the selected operation-level dependencies, together with $\so$, $\app$, and the derived $\rw$ edges, must form an acyclic 
    operation
 graph.
This rules out resolutions that assign source
appends to read prefixes in a manner inconsistent with the ordering of operations. 
Second, after lifting the selected dependencies to transactions,
 the resulting transaction-level dependency graph must also be acyclic.
This graph is obtained by collapsing the operation-level dependency graph onto original transactions: 
auxiliary unfolded reads are ignored, and a transaction-level dependency $T \to T'$ is introduced whenever the resolution contains a dependency  from an original operation in $T$ to an original operation in $T'$.

%A valid resolution must also be acyclic at both levels. 
%First, the selected operation-level dependencies, 
%together with the operation-level $\so$ order, the program order within transactions, 
%the $\app$ edges, and the derived operation-level $\rw$ edges,  
%must form an acyclic operation graph. 
%This rules out resolutions that assign sources to read prefixes in a way that is
%inconsistent with the order of operations. 
%Second, after lifting the selected operation-level dependencies to transactions, 
%the resulting transaction-level dependency graph must also be acyclic. 
%This graph is obtained by collapsing only original operations 
%in the operation-level graph. 
%That is, auxiliary unfolded reads are ignored,
%and we add a transaction-level edge $T \to T'$ 
%whenever the resolution contains a dependency edge 
%from an original operation in $T$ to an original
%operation in $T'$. 

\begin{theorem} \label{thm:list-ser}
	A list-append history satisfies 
	serializability
	 iff there exists a resolution of the operation-level choice sets  
	 \begin{itemize}[leftmargin=20pt]
\item[(i)]
satisfying the append-order condition; and
\item[(ii)]
yielding acyclic dependency graphs at both the operation and transaction levels.\footnote{A proof is provided in Appendix~\ref{app:proof}.}
	 \end{itemize}

%  A list-append history satisfies serializability
%   iff 
%  there exists a resolution of the operation-level constraints such that 
%  (i) the append-order condition is satisfied; 
%   and 
%  (ii)
%  the resulting dependency graphs are acyclic at both the operation and transaction levels.
\end{theorem}

Figure~\ref{fig:list-dup-solution} shows a valid resolution of the
history in Figure~\ref{fig:list-dup-constraints}. 
%This resolution satisfies the operation-level adjacency condition. 
The selected $\wr$
edges identify the source append for each successive list extension:
the initial append $\appendevent(x,\emptypost)$ accounts for \mylist{[]},
$\appendevent(x,1)$ in $T_1$ accounts for \mylist{[1]},
$\appendevent(x,2)$ in $T_2$ accounts for \mylist{[1,2]},
and the later append $\appendevent(x,1)$ in $T_2$ accounts for
\mylist{[1,2,1]}. The selected $\ww$ edges order these appends
 as
required by the $\app$ edges, and each consecutive pair is adjacent in
the $\ww(x)$ order. Thus, the resolution explains the observed list at
the operation level. 
%Due to space limitations, 
%we defer 

Figure~\ref{fig:list-dup-solution-txn} shows the corresponding Adya dependency graph, 
which is also acyclic.
This transaction-level graph  is obtained by excluding the auxiliary unfolded reads
and lifting the dependencies between the remaining operations
to their enclosing transactions. 
For example, the chain 
$T_{\bot}\!:\!\appendevent(x,\emptypost)
\rel{\ww(x)}
T_1\!:\!\appendevent(x,1)
\rel{\ww(x)}
T_2\!:\!\appendevent(x,2)$
induces
$T_{\bot} \rel{\WW(x)} T_1 \rel{\WW(x)} T_2$;
 the edge 
$T_2\!:\!\appendevent(x,1) \rel{\wr(x)} T_3\!:\!\readevent(x,[1,2,1])$
induces
$T_2 \rel{\WR(x)} T_3$.

\section{The Algorithm}
Algorithm~\ref{alg:check-list} summarizes
our procedure
\textsc{Verify} 
 for verifying  list-append histories
with duplicate values.
Given a history $\H$, 
it
first invokes \textsc{Unfold} to obtain the operation set $\operations$, the operation-level session order $\so$, the program order $\po$, and the append order $\app$ (Line~\ref{line:unfold}).
It then constructs 
the operation-level choice sets $\wwopcons$ and $\wropcons$ 
for possible $\ww$ and $\wr$ dependencies (Line~\ref{line:construct-constraints}).

Next, the algorithm enumerates all resolutions $(\ww,\wr)$ of these choice sets (Line~\ref{line:enumerate-resolutions}).
For each resolution, it derives the corresponding operation-level $\rw$ edges (Line~\ref{line:derive-rw}) and checks the two conditions in Theorem~\ref{thm:list-ser} (Line~\ref{line:test-condition}): 
(i) the selected $\wr$ and $\ww$ edges satisfy the append-order condition, as checked by \textsc{CheckAOCond}; and
(ii) the resulting dependency graphs are acyclic at both the operation and transaction levels, as checked by \textsc{CheckDepAcyc}.
If a resolution satisfies both conditions, the history is accepted as
serializable (Line~\ref{line:return-true}); 
otherwise, it is rejected (Line~\ref{line:return-false}).

%\inlsec{Complexity Analysis}
The worst-case complexity of the algorithm is $n^{O(n^2)}$, 
with $n$  the number of operations in $\H$
(see Appendix~\ref{app:complexity}).
Yet, this bound is reached only in the extreme case where all transactions append the same value to the same key.
Such workloads are unlikely 
to arise 
under the randomized workload generation commonly used in practice (e.g., Jepsen).
As we will see next, our tool
 remains practical even under highly skewed workloads with a high proportion of duplicate values.

\begin{algorithm}[t]
  \small
  \caption{Verifying \duplicateval{} list-append histories}
  \label{alg:check-list}
  \begin{algorithmic}[1]
    \Function{Verify}{$\H$} 
      \label{procedure:check}
      \State $(\operations, \so, \po, \app) \gets \Call{Unfold}{\H}$
        \label{line:unfold}
      \State $(\wwopcons, \wropcons) \gets \Call{ConstructChoices}{\operations}$
        \label{line:construct-constraints}
      \For {each resolution $(\ww, \wr)$ of $(\wwopcons, \wropcons)$} \label{line:enumerate-resolutions}
        \State {$\rw \gets \Call{DeriveRW}{\ww, \wr}$} \label{line:derive-rw}
        \If {%
            $\begin{aligned}[t]
              &\Call{CheckAOCond}{\ww, \wr, \app} \land{} \\[-3pt]
              &\Call{CheckDepAcyc}{\so, \po, \app, \ww, \wr, \rw}
            \end{aligned}$}
            \label{line:test-condition}
          \State{\Return $\true$} \label{line:return-true}
        \EndIf
      \EndFor
      \State {\Return $\false$} \label{line:return-false}
    \EndFunction
  \end{algorithmic}
\end{algorithm}
\vspace{-2ex}

\section{Experiments}

We have implemented our algorithm in a prototype checker,  called
 \ourtool, 
 with approximately 3K lines of C++ code.
It utilizes \emph{hyper-polygraphs}~\cite{veristrong}
 to
encode
both operation-level and transaction-level 
 dependency constraints, 
 which are then solved using 
an SMT solver~\cite{veristrong} 
optimized for  checking strong isolation levels like serializability.

We conduct a preliminary evaluation of \ourtool along two dimensions:
 its effectiveness in detecting isolation bugs and its checking performance.

\inlsec{Rediscovering  Duplicate-Value Bugs}
%We apply \ourtool{} to test both MySQL and MariaDB. 
%Using \duplicateval{} list-append workloads, 
%\ourtool{} successfully reproduces known consistency bugs in MySQL 8.0 and MariaDB 11.5, 
%which were originally reported using rw-register workloads~\cite{MySQL-Bug,MariaDB-Bug}. 
%This result suggests that our approach has the potential to discover real-world anomalies 
%whose detection requires resolving ambiguity caused by duplicate values.
Using \duplicateval{} list-append workloads,
 \ourtool 
 successfully
  rediscovers two serializability violations in MySQL (v8.0) and MariaDB (v11.5)
that
   were originally reported 
   through SQL test cases~\cite{MySQL-Bug,MariaDB-Bug}. 
Both violations require duplicate values to manifest 
and  therefore lie beyond the scope of most existing isolation validators, 
 including Elle.
 These findings also highlight the potential of duplicate-value list-append workloads
 to expose subtle isolation bugs in real database systems.

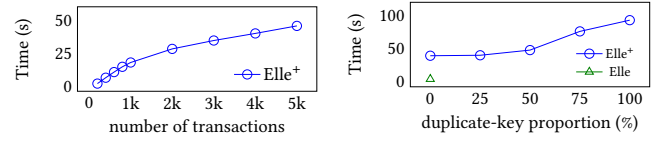
\begin{figure}[t]
	\centering
	\begin{minipage}[b]{0.48\linewidth}
		% txns.tex

\pgfplotsset{height=100pt, width=200pt}
\pgfplotsset{tick style={draw=none}}

  \centering
  \begin{scaletikzpicturetowidth}{\textwidth}
		\begin{tikzpicture}[scale=\tikzscale]
			\begin{axis}[
        font=\Large,
				xlabel={number of transactions},
        xtick={0, 50, 100, 150, 200, 250}, 
        xticklabels={0, 1k, 2k, 3k, 4k, 5k}, 
				ylabel={Time (s)},
        ytick={0,25,50},
        ymax=60,
				legend pos=south east,
legend style={
	draw=none
}		
				]
				\addplot[color=blue,mark=o,mark size=3pt] table [x=param, y=ours-list, col sep=comma] {figs/eval/various-list/data-w-mem/runtime-txn.csv};
				% \addplot[color=brown,mark=square,mark size=3pt] table [x=param, y=ours-rw, col sep=comma] {figs/eval/various-list/data-w-mem/runtime-txn.csv};
        \legend{\ourtool}
			\end{axis}
		\end{tikzpicture}
  \end{scaletikzpicturetowidth}
  %						\captionsetup{skip=5pt}
 % \caption{Runtime of \ourtool as the number of transactions increases.}
  % #sessions: 20
  % #txns/sessions: x-axis/20
  % #ops/txn: 20
  % #keys: 5k
  % 50% read + 50% write
  % 50% keys duplicate value + 50% keys unique value
  % value distribution: zipfian(theta = 0.5, N = 100)
%  \label{fig:performance}
	\end{minipage}
	\hfill
	\begin{minipage}[b]{0.48\linewidth}
		% list-rw.tex

\pgfplotsset{height=100pt, width=200pt}
\pgfplotsset{tick style={draw=none}}

% \begin{figure}[h]
	\centering
	\begin{scaletikzpicturetowidth}{\textwidth}
		\begin{tikzpicture}[scale=\tikzscale]
			\begin{axis}[
				font=\Large,
				title={},
				xlabel={duplicate-key proportion (\%)},
				xtick={0,0.25,0.5,0.75,1},
				xticklabels={0,25,50,75,100},
				xticklabel style={/pgf/number format/fixed},
				scaled x ticks=false,
				ylabel={Time (s)},
				ymax=120,
        % ymode=log,
				cycle multiindex* list={
					color       \nextlist
					mark list*  \nextlist
				},
				legend pos=south east,
    legend style={
	draw=none,
	    font=\small
}				
			]
				% \addplot[color=blue,mark=o,mark size=3pt] table [x=param, y=ours-rw, col sep=comma] {figs/eval/list-rw/data-no-single-write.csv};
				\addplot[color=blue,mark=o,mark size=3pt] table [x=param, y=ours-list, col sep=comma] {figs/eval/list-rw/data-no-single-write.csv};
				\addplot[color=green!50!black,mark=triangle,mark size=3pt] table [x=param, y=elle, col sep=comma] {figs/eval/list-rw/data-no-single-write.csv};
				\legend{\ourtool,Elle}
			\end{axis}
		\end{tikzpicture}
	\end{scaletikzpicturetowidth}
		%					\captionsetup{skip=5pt}
%  \caption{Runtime under varying duplicate-key proportions.}
	% #sessions: 100
	% #txns/session: 100
	% #ops/txn: 8
	% #keys: 5k
	% 50% read + 50% write
	% duplicate key proportion: x-axis
	% value distribution: zipfian(theta = 1.5, N = 100)
%	\label{fig:compare-with-elle}
% \end{figure}
	\end{minipage}
	  						\captionsetup{skip=5pt}
	\caption{Runtime  under varying number of transactions (left) and duplicate-key proportions (right).}
	\label{fig:exp}
  \vspace{-2ex}
\end{figure}

\inlsec{Performance Evaluation} 
To obtain \duplicateval{} histories,
we extend the workload generator of~\cite{Complexity:OOPSLA2019},
 integrate it with Jepsen,
 and
run the generated workloads against 
a PostgreSQL (v15) instance 
under serializability. 
The extended generator
allows us to control the fraction of keys 
receiving 
duplicate-value appends
and
uses a Zipfian key-access distribution to emulate high-contention workloads.\footnote{All parameters and their  values are summarized in Appendix~\ref{app:para}.}
Our experiments run on a local machine 
equipped with an Intel 13th Gen i5 CPU and 32GB of RAM. 

Figure~\ref{fig:exp}  (left)
 shows
 that
  the checking time of Elle$^+$ 
 grows steadily with workload size,  
 reaching approximately 50s for histories containing 5K transactions 
 (and 100K operations).

% Figures~\ref{fig:performance} and~\ref{fig:compare-with-elle} report the
% runtime of \ourtool{} under different workloads. 
Figure~\ref{fig:exp} (right) shows the runtime as the fraction of keys receiving duplicate-value appends increases.
As expected, Elle  outperforms \ourtool
 in the \uniqueval{} setting, i.e., when the proportion is zero.
The performance gap stems mainly from 
the need to solve additional constraints introduced at the operation level. 
Yet, 
\ourtool manages to 
verify histories containing 10K transactions and 80K operations
in under 100s, even when all 5K keys receive duplicate-value appends.

%Figure~\ref{fig:performance} shows how the runtime changes as the number of transactions increases, 
%while Figure~\ref{fig:compare-with-elle} shows the runtime trend as the proportion
%of duplicate keys increases. 
%When this proportion is zero, the generated histories satisfy the \uniqueval{} assumption and 
%can therefore be checked by Elle, enabling a direct comparison. 
%In this \uniqueval{} setting, 
%Elle shows a clear performance advantage. 
%The overhead of \ourtool{} mainly comes from its
%use of SMT solving and its fine-grained operation-level modeling.

%\input{sec/related}

%\input{sec/discussion}

% conclusion.tex

\section{Future Work}
\label{sec:concl}

% \zhiheng{discuss ``can our approach extend to other isolation levels?'' For \si, \yes; 
% for other formalisms, it requires adaptation of dependency graphs}

%We have presented Elle$^+$,
% an extension of Elle that enables sound and complete isolation checking for workloads containing duplicate values.
%Our evaluation have demonstrated that Elle$^+$ can successfully reproduce realistic serializability violations that require duplicate values to manifest, 
%while maintaining promising checking performance.

Building on
 the promising preliminary results of Elle$^+$,
future work includes 
(i) leveraging Jepsen's fault injection  for large-scale testing of distributed database systems,
(ii) exploring additional optimizations to improve the scalability of Elle$^+$,
and
(iii) extending it  to support other isolation levels~\cite{noc-noc,si}.

\clearpage
\balance
\bibliographystyle{ACM-Reference-Format}
\bibliography{ref}

@article{PolySI:VLDB2023,
	author = {Kaile Huang and Si Liu and Zhenge Chen and Hengfeng Wei and David A. Basin and Haixiang Li and Anqun Pan},
	title = {Efficient Black-box Checking of Snapshot Isolation in Databases},
	journal = {Proc. {VLDB} Endow.},
	volume = {16},
	number = {6},
	pages = {1264--1276},
	year = {2023}
}

@inproceedings{Viper:EuroSys2023,
	author = {Zhang, Jian and Ji, Ye and Mu, Shuai and Tan, Cheng},
	title = {Viper: A Fast Snapshot Isolation Checker},
	year = {2023},
	publisher = {ACM},
	booktitle = {EuroSys 2023},
	pages = {654–671}
}

@inproceedings{Cobra:OSDI2020,
	author       = {Cheng Tan and
	Changgeng Zhao and
	Shuai Mu and
	Michael Walfish},
	title        = {Cobra: Making Transactional Key-Value Stores Verifiably Serializable},
	booktitle    = {{OSDI} 2020},
	pages        = {63--80},
	publisher    = {{USENIX} Association},
	year         = {2020}
}

@article{Plume:OOPSLA2024,
	author = {Liu, Si and Gu, Long and Wei, Hengfeng and Basin, David},
	title = {Plume: Efficient and Complete Black-Box Checking of Weak Isolation Levels},
	year = {2024},
	issue_date = {October 2024},
	publisher = {Association for Computing Machinery},
	address = {New York, NY, USA},
	volume = {8},
	number = {OOPSLA2},
	url = {https://doi.org/10.1145/3689742},
	doi = {10.1145/3689742},
	journal = {Proc. ACM Program. Lang.},
	month = oct,
	articleno = {302},
	numpages = {29},
}

@article{Elle:VLDB2020,
	author = {Kingsbury, Kyle and Alvaro, Peter},
	title = {Elle: Inferring Isolation Anomalies from Experimental Observations},
	year = {2020},
	issue_date = {November 2020},
	publisher = {VLDB Endowment},
	volume = {14},
	number = {3},
	issn = {2150-8097},
	journal = {Proc. VLDB Endow.},
	pages = {268-280},
	numpages = {13}
}

@article{Complexity:OOPSLA2019,
	author = {Biswas, Ranadeep and Enea, Constantin},
	title = {On the Complexity of Checking Transactional Consistency},
	year = {2019},
	issue_date = {October 2019},
	publisher = {ACM},
	volume = {3},
	number = {OOPSLA},
	journal = {Proc. ACM Program. Lang.},
	articleno = {165},
}

@misc{MariaDB-Bug,
	author = {MariaDB-\#26642},
	title = {Weird {SELECT} view when a record is modified to the same value by two transactions},
	year = {Accessed in May, 2026},
	howpublished = {\url{https://jira.mariadb.org/browse/MDEV-26642}}
}

@misc{MySQL-Bug,
	author = {MySQL-\#100328},
	title = {Inconsistent behavior with isolation levels when binlog enabled},
	year = {Accessed in May, 2026},
	howpublished = {\url{https://bugs.mysql.com/bug.php?id=100328}}
}

@article{si,
	title={A critique of {ANSI} {SQL} isolation levels},
	author={Berenson, Hal and Bernstein, Phil and Gray, Jim and Melton, Jim and O'Neil, Elizabeth and O'Neil, Patrick},
	journal={ACM SIGMOD Record},
	volume={24},
	number={2},
	pages={1--10},
	year={1995},
	publisher={ACM}
}

@phdthesis{Adya:PhDThesis1999,
	author = {Adya, Atul},
	title = {Weak Consistency: A Generalized Theory and Optimistic Implementations for Distributed Transactions},
	year = {1999},
	publisher = {Massachusetts Institute of Technology},
	address = {USA}
}

@misc{Jepsen,
	author = {{Jepsen}},
	year = {Accessed in June, 2026},
	note = {\url{https://jepsen.io}}
}

@inproceedings{Framework:CONCUR2015,
	author       = {Andrea Cerone and
	Giovanni Bernardi and
	Alexey Gotsman},
	title        = {A Framework for Transactional Consistency Models with Atomic Visibility},
	booktitle    = {{CONCUR} 2015},
	series       = {LIPIcs},
	volume       = {42},
	pages        = {58--71},
	publisher    = {Schloss Dagstuhl - Leibniz-Zentrum f{\"{u}}r Informatik},
	year         = {2015}
}

@inproceedings{MiniTransactions:ICDE2025,
	author       = {Hengfeng Wei and
	Jiang Xiao and
	Na Yang and
	Si Liu and
	Zijing Yin and
	Yuxing Chen and
	Anqun Pan},
	title        = {Boosting End‑to‑End Database Isolation Checking via Mini‑Transactions},
	booktitle    = {ICDE 2025},
	year = {2025},
	pages = {3998-4010},
	publisher = {IEEE Computer Society},
	month =May
}

@article{AWDIT:PLDI2025,
	author = {M\o{}ldrup, Lasse and Pavlogiannis, Andreas},
	title = {AWDIT: An Optimal Weak Database Isolation Tester},
	year = {2025},
	issue_date = {June 2025},
	publisher = {ACM},
	volume = {9},
	number = {PLDI},
	journal = {Proc. ACM Program. Lang.},
	month = jun,
	articleno = {209},
	numpages = {25}
}

@article{noc-noc,
	author = {Liu, Si and Multazzu, Luca and Wei, Hengfeng and Basin, David A.},
	title = {NOC-NOC: Towards Performance-optimal Distributed Transactions},
	year = {2024},
	issue_date = {February 2024},
	publisher = {ACM},
	volume = {2},
	number = {1},
	journal = {Proc. ACM Manag. Data},
	month = mar,
	articleno = {9},
	numpages = {25}
}

@inproceedings{DBLP:conf/sigmod/WarszawskiB17,
	author       = {Todd Warszawski and
	Peter Bailis},
	title        = {ACIDRain: Concurrency-Related Attacks on Database-Backed Web Applications},
	booktitle    = { {SIGMOD} 2017},
	pages        = {5--20},
	publisher    = {{ACM}},
	year         = {2017}
}

@inproceedings{10.1145/3597503.3639207,
	author = {Cui, Ziyu and Dou, Wensheng and Gao, Yu and Wang, Dong and Song, Jiansen and Zheng, Yingying and Wang, Tao and Yang, Rui and Xu, Kang and Hu, Yixin and Wei, Jun and Huang, Tao},
	title = {Understanding Transaction Bugs in Database Systems},
	year = {2024},
	isbn = {9798400702174},
	publisher = {ACM},
	booktitle = {ICSE '24},
	articleno = {163},
	numpages = {13}
}

@article{veristrong,
	author = {Cai, Zhiheng and Liu, Si and Wei, Hengfeng and Chen, Yuxing and Pan, Anqun},
	title = {Fast Verification of Strong Database Isolation},
	year = {2025},
	issue_date = {December 2025},
	publisher = {VLDB Endowment},
	volume = {19},
	number = {4},
	issn = {2150-8097},
	journal = {Proc. VLDB Endow.},
	month = dec,
	pages = {563–575},
	numpages = {13}
}

@inproceedings{leopard,
	author       = {Keqiang Li and
	Siyang Weng and
	Peiyuan Liu and
	Lyu Ni and
	Chengcheng Yang and
	Rong Zhang and
	Xuan Zhou and
	Jianghang Lou and
	Gui Huang and
	Weining Qian and
	Aoying Zhou},
	title        = {Leopard: {A} Black-Box Approach for Efficiently Verifying Various
	Isolation Levels},
	booktitle    = {{ICDE} 2023},
	pages        = {722--735},
	publisher    = {{IEEE}},
	year         = {2023}
}

\clearpage
\appendix
%\input{sec/app-graph}

% app-list-proof-sketch.tex 

\section{Proof of Theorem~\ref{thm:list-ser}}
\label{app:proof}
For a total order $R$, we write $a \relimm{R} b$ if
$a \rel{R} b$ and there is no element $c$ such that
$a \rel{R} c \rel{R} b$.

We use the $\vis, \ar$ characterization of \ser{}~\cite{Framework:CONCUR2015}
%~\zhiheng{be careful that CONCUR2015 may not appear in the main text}
in terms of an abstract execution $\ae=(\T,\SO,\vis,\ar)$ satisfying
\sessionaxiom{}, \intextaxiom{}, and \totalvis{}.
Since \totalvis{} requires $\vis=\ar$, the arbitration order induces a
total operation-level order, denoted by $\gar$, by extending $\ar$ with
program order inside transactions. Similarly, $\vis$ induces an
operation-level visibility relation $\gvis$.

The only list-specific semantic condition used below is the following.
For a read operation
$r \triangleq \readevent(x,[v_1,\ldots,v_n])$, the visible append
operations on key $x$ must be exactly
$w_0,w_1,\ldots,w_n$ in $\gar$ order, where
$w_0=\appendevent(x,\emptypost)$ and
$w_i=\appendevent(x,v_i)$ for every $1\leq i\leq n$.

\begin{proofsketch}
We prove Theorem~\ref{thm:list-ser} in both directions.

(``$\Leftarrow$'')
Suppose there exists a valid resolution, and let $\G'$ and $\G'_{\T}$
be the resulting operation-level and transaction-level dependency
graphs. By validity, both graphs are acyclic. We take $\ar$ to be a
topological order of $\G'_{\T}$ and set $\vis=\ar$. Then
\sessionaxiom{} and \totalvis{} hold immediately, since $\SO$ is
included in $\G'_{\T}$ and $\vis=\ar$ is total.

It remains to check the list semantics required by \intextaxiom{}.
Consider an arbitrary read
$r \triangleq \readevent(x,[v_1,\ldots,v_n])$ and its unfolded reads
\[
  r'_0 \rel{\app(x)} r'_1 \rel{\app(x)} \cdots
  \rel{\app(x)} r'_{n-1} \rel{\app(x)} r .
\]
Let $w_0,w_1,\ldots,w_n$ be the append operations selected by the
$\wr(x)$ edges for these reads. By construction of $\wr$ constraints,
these operations append exactly
$\emptypost,v_1,\ldots,v_n$, respectively. Moreover, the
append-order condition ensures
\[
  w_0 \relimm{\ww(x)} w_1
  \relimm{\ww(x)} \cdots
  \relimm{\ww(x)} w_n .
\]
Thus these append operations explain the successive extensions of the
list read by $r$.

There cannot be any additional visible append operation on key $x$
before $r$. Indeed, because $w_0,\ldots,w_n$ are adjacent in the
$\ww(x)$ order, any such extra append $w'$ would have to satisfy
$w_n \rel{\ww(x)} w'$. Together with
$w_n \rel{\wr(x)} r$, the $\rw$ derivation rule gives
$r \rel{\rw(x)} w'$, forcing $r$ to precede $w'$ in the dependency
graph, contradicting the assumption that $w'$ is visible to $r$ under
the topological order. Hence the visible appends of $r$ are exactly
$w_0,\ldots,w_n$, so \intextaxiom{} holds. Therefore the history
satisfies \ser{}.

(``$\Rightarrow$'')
Conversely, suppose the history satisfies \ser{}, witnessed by an
abstract execution $\ae=(\T,\SO,\vis,\ar)$. Since \totalvis{} holds, we
have $\vis=\ar$. We construct a resolution from this abstract execution.

First, order append operations on each key according to $\gar$ and use
this order to select the $\ww$ edges. For each read prefix introduced by
unfolding, select as its $\wr$ source the maximal visible append on the
same key under $\gar$. The list semantics of \intextaxiom{} guarantees
that this source has the required value, so the selected $\wr$ edges are
among the alternatives generated by the constraint construction. The
$\rw$ edges are then derived from the selected $\ww$ and $\wr$ edges.

This resolution satisfies the append-order condition. Indeed, two
adjacent unfolded reads correspond to two consecutive prefixes of the
same observed list. By \intextaxiom{}, these prefixes must be produced
by consecutive visible append operations on the same key. Therefore, if
their selected sources are $w$ and $w'$, then
$w \rel{\ww(x)} w'$, and there is no append $w''$ such that
$w \rel{\ww(x)} w'' \rel{\ww(x)} w'$.

It remains to argue acyclicity. First ignore the auxiliary unfolded read
operations and consider the graph $\G'_1$ over the original operations.
The $\po$, $\so$, $\ww$, and $\wr$ edges are consistent with $\gar$ by
construction. A derived edge $r \rel{\rw(x)} w'$ can only arise from
$w \rel{\wr(x)} r$ and $w \rel{\ww(x)} w'$. If $w'$ preceded $r$ in
$\gar$, then $w'$ would be visible to $r$, contradicting the maximality
of $w$ as the visible append source of $r$. Hence $\rw$ edges are also
consistent with $\gar$, and $\G'_1$ is acyclic. The lifted
transaction-level graph is acyclic for the same reason, since all lifted
edges are consistent with $\ar$.

Finally, add the unfolded reads and the $\app$ edges to obtain the full
operation-level graph $\G'_2$. Suppose, for contradiction, that $\G'_2$
contains a cycle. If the cycle contains no unfolded read, it is already
a cycle in $\G'_1$, impossible. 
It remains to show that the unfolded reads do not introduce new cycles.
We prove that any cycle in $\G'_2$ would imply a cycle in $\G'_1$:
cycles passing through unfolded reads can be eliminated using the
append-order condition. Since $\G'_1$ is acyclic, $\G'_2$ is
acyclic as well.

Hence the constructed resolution satisfies the append-order
condition and both acyclicity requirements, so it is a valid resolution.
This completes the proof sketch.
\end{proofsketch}

\section{Complexity Analysis}
\label{app:complexity}
Let $n$ be the number of operations in the input history. 
The running time of Algorithm~\ref{alg:check-list} is dominated by enumerating
possible resolutions of the uncertain $\ww$ and $\wr$ dependencies.
After read unfolding, there can be $O(n^2)$ $\ww$ choice sets with two
alternatives each, and $O(n^2)$ $\wr$ choice sets with at most $O(n)$
alternatives each. 
Thus, the number of resolutions is bounded by $2^{O(n^2)} n^{O(n^2)}$. 
For each resolution, the validity checks,
including deriving $\rw$, checking append-order condition, and checking
acyclicity at both levels, take polynomial time, say $O(n^4)$ under a
straightforward implementation. Hence the overall worst-case complexity
is $n^{O(n^2)}$. 
Yet, this upper bound is reached only in the extreme case 
where all transactions append the same value to the same key. 
Real-world workloads are unlikely to exhibit such a worst-case pattern, 
and thus typically induce a much smaller search space.

\section{Workload Generator Parameters}
\label{app:para}

In Figure~\ref{fig:exp}~(left), we vary the total number of transactions.
Each generated history contains 20 sessions, 
and each transaction contains 20 operations. 
The key space contains 5,000 keys, 
and the workload consists of 50\% reads and 50\% appends. 
Half of the keys allow duplicate append values, while the other half use unique append values. 
For each key, append values are sampled from a Zipfian distribution 
with skew parameter $\theta = 0.5$ over a domain of size $N = 100$.

In Figure~\ref{fig:exp}~(right), 
we vary the proportion of keys that allow duplicate append values. 
Each generated history contains 100 sessions, 
each with 100 transactions, and each transaction contains 8 operations. 
The key space again contains 5,000 keys, 
and the workload consists of 50\% reads and 50\% appends. 
Append values are sampled from a Zipfian distribution 
with skew parameter $\theta = 1.5$ over a domain of size $N = 100$.

\end{document}